\documentstyle[11pt,mrs2001,epsfig]{article} 
\begin{document}
\title{Constraining cosmological models using arc statistics in future
SZ cluster surveys}

\author{M. MENEGHETTI$^{1,2}$, M. BARTELMANN$^2$, L. MOSCARDINI$^1$}
\affil{$^1$Dipartimento di Astronomia, Universit\`a di Padova, vicolo
dell'Osservatorio 2, I--35122 Padova, Italy}
\affil{$^2$Max-Planck-Institut f\"ur Astrophysik, P.O.~Box 1317,
D--85741 Garching, Germany}

\begin{abstract}
Upcoming wide-area surveys in the submillimetre regime will allow the
construction of complete galaxy cluster samples through their thermal
Sunyaev-Zel'dovich effect. We propose an analytic method to predict
the number of gravitationally lensed giant arcs produced by the
cluster sample expected to be detectable for the {\em Planck\/}
satellite. The statistics of lensed arcs can be used to constrain
cosmological parameters. This computation implies the choice of an
analytic cluster model which has to reproduce the lensing properties
of realistic galaxy clusters. Comparing the results of analytic
computations and numerical ray-tracing simulations, we show that
spherical models with the Navarro-Frenk-White density profile fail by
far to produce as many giant arcs as realistic galaxy clusters. This
confirms the great importance of asymmetries and substructures in the
lensing matter distribution. By elliptically distorting the lensing
potential of spherical NFW density profiles, we show that the
discrepancy between analytic and numerical computations can almost
completely be removed.
\end{abstract}

\section{Introduction}

Several studies demonstrated that arc statistics is a very sensitive
tool to constrain cosmological models. For example, using ray-tracing
simulations and numerically simulated galaxy clusters, Bartelmann et
al.~\cite{bartelmann98} showed that the expected number of giant arcs
(defined as arcs with length-to-width ratio larger than 10 and $B$
magnitude less than 21.5) changes by orders of magnitude from low
density to high density universes, depending also on the value of the
cosmological constant. This result, with the exception of the
dependency on the cosmological constant, was confirmed by several
other authors \cite{cooray99,kauffmann00}, who used spherical analytic
and somewhat idealistic models to describe the cluster lenses and
predict the number of giant arcs as a function of the cosmological
parameters. Therefore, the comparison of the expectation from such
theoretical studies to the observations of strong lensing events in
galaxy clusters can be used to constrain the cosmological model.

In order to make such a comparison, a complete sample of galaxy
clusters could be selected through the thermal Sunyaev-Zel'dovich
effect (hereafter SZ effect). This will be possible in the near future
thanks to the upcoming full-sky microwave surveys, like that planned
for the {\em Planck\/} satellite mission.  Although other selection
methods could be used for this purpose (i.e.~through the X-ray
emission from the intra-cluster gas), the SZ selection criterion is
much less sensitive to biases due to the dynamical processes in the
cluster centre, like cooling flows or merging events.

\section{Lensing Sunyaev-Zel'dovich clusters}

Upcoming full-sky microwave surveys will detect of order of $10^4$
galaxy clusters through their thermal SZ effect. This effect is
determined by the Compton-$y$ parameter
\begin{equation}
  y(\vec{\theta})=\frac{k}{m_{\rm e}c^2}\sigma_{\rm T}\int{\rm d}l\,
  T(\vec{\theta},l)\,n_{\rm e}(\vec{\theta},l)\;,
\end{equation}
where $T$ is the temperature of the ionized intra-cluster gas and
$n_{\rm e}$ is the electron number density. The clusters detectable
through their thermal SZ effect are those for which
\begin{equation}	
  Y\equiv\int{\rm d}^2\vec{\theta}y(\vec{\theta}) \ge Y_{\rm min}\;,
\end{equation}
where the integral is performed over the source area. The lower limit
$Y_{\rm min}$ depends on the angular resolution and the temperature
sensitivity of the detector. For example, for the {\em Planck\/}
satellite, $Y_{\rm min}$ is expected to be $\approx 10^{-4}$
arcmin$^2$.

As explained in more detail by Bartelmann \cite{bartelmann00}, many SZ
clusters will also be efficient weak gravitational lenses. Their weak
lensing effects can be quantified using the {\em aperture mass\/}
\cite{schneider96},
\begin{equation}
  M_{\rm ap}(\theta)= \int \mbox{d}^2 \vec{\vartheta}
  \kappa(\vec{\vartheta})U(|\vec{\vartheta}|)\;,
\end{equation}
which is the integral within a circular aperture with radius $\theta$
over the convergence $\kappa$, weighted by a suitable filter function
satisfying the condition
\begin{equation}
  \int_0^{\theta} \mbox{d}^2 \vartheta \vartheta U(\vartheta) = 0 \;.
\end{equation}

The dispersion of $M_{\rm ap}$ due to the finite number density
$n_{\rm g}$ of randomly distributed background galaxies and their
intrinsic ellipticity dispersion $\sigma_\epsilon$ was computed by
Schneider \cite{schneider96},
\begin{equation}
  \sigma_M(\theta)=0.016\,
  \left(\frac{n_{\rm g}}{30\,\mbox{arcmin}^2}\right)^{-1/2}
  \left(\frac{\sigma_\epsilon}{0.2}\right)\,
  \left(\frac{\theta}{1\,\mbox{arcmin}}\right)^{-1}\;.
\end{equation}

Clusters produce a significant weak-lensing effect if the
signal-to-noise ratio $S=M_{\rm ap}(\theta)/\sigma_M({\theta})$ is
larger than a minimal value $S_{\rm min}$.  Assuming $S_{\rm min}=5$,
Bartelmann \cite{bartelmann00} estimated that more than $70\%$ of the
original SZ cluster sample expected from {\em Planck\/} will be
efficient at least for weak lensing, and many of them will also show
strong lensing features.

\section{Number of arcs}

The number of arcs which is expected to be observed in the selected
cluster sample can be estimated using the following information:

\begin{itemize}

\item both conditions, that clusters must be detectable through the
thermal SZ effect and be efficient weak gravitational lenses, give a
minimal cluster mass $M_{\rm min}(z_{\rm L})$ at any given redshift
$z_{\rm L}$ \cite{bartelmann00};

\item the number density of dark matter haloes $n_{\rm PS}(M,z_{\rm
L})$ can be modeled using the Press \& Schechter \cite{press74} mass
function or its generalizations, like the Sheth \& Tormen
\cite{sheth99} mass function;

\item the lensing cross section for arcs with a minimal
length-to-width ratio $L/W$ (approximately corresponding to a minimal
magnification $\mu$) $\sigma(L/W,\mu;M,z_{\rm L},z_{\rm S})$ can be
computed for any galaxy cluster with mass $M$ at redshift $z_{\rm L}$
and for any source redshift $z_{\rm S}$ once the analytic model
describing the cluster lens is chosen;

\item the source galaxy redshift distribution can be modeled as 
\begin{equation}
  p(z_{\rm S}) \propto z_{\rm S}^2 \exp(-z_{\rm S}^\beta)\;,
\end{equation}
where $\beta=1.5$ \cite{smail95};

\item finally, the differential number density of sources with a given
observed flux $F$ can be assumed to be a power law:
\begin{equation}
  \frac{\mbox{d}n_{\rm S}}{\mbox{d}F} \propto F^{-\alpha}\;,
\end{equation}
where $\alpha\approx1.8$ in the R-band \cite{smail95}.

\end{itemize}

Using these ingredients, the optical depth for the formation of arcs
with length-to-width ratio larger than $L/W$ is given by
\begin{equation}
  \tau(L/W,\mu;z_{\rm S}) = \frac{1}{4\pi D_{\rm S}^2}\,
  \int_0^{z_{\rm S}} \int_{M_{\rm min}}^{\infty}
  \sigma\,\mbox{d}z_{\rm L}\,
  \left|\frac{\mbox{d}V}{\mbox{d}z_{\rm L}}\right|\,
  n_{\rm PS}(M,z_{\rm L})(1+z_{\rm L})^3\mbox{d}M\;,
\end{equation}
where $D_{\rm S}$ is angular diameter distance to the sources at
redshift $z_{\rm S}$, and $\mbox{d}V/\mbox{d}z$ is the proper volume
per unit redshift.

Finally, the number of arcs with $L/W \ge (L/W)_{\rm min}$ and
magnitude $R\le R_{\rm max}$ (corresponding to the minimal observed
flux $F_0$) is given by
\begin{equation}
  N=\int_0^\infty \int_{\mu_{\rm min}}^{\infty}
  \tau(L/W,\mu;z_{\rm S})\,\left(\int_{F_0}^\infty
  \frac{\mbox{d}n_{\rm S}}{\mbox{d}(F/\mu)}
  \frac{\mbox{d}F}{\mu}\right)\,
  p(z_{\rm S})\mbox{d}\mu\mbox{d}z_{\rm S}\;.
\end{equation}

\section{Which analytic model?}

The crucial point in this computation is the choice of an analytic
model for describing the lenses. Several authors
\cite{cooray99,kauffmann00} modeled them as isothermal spheres in
order to measure their efficiency to produce giant arcs. In this
study, we use the much more realistic Navarro-Frenk-White (hereafter
NFW) density profile, and we compare the strong lensing properties of
this analytic model with those of realistic galaxy clusters obtained
from $N$-body simulations. The NFW profile has strong-lensing
properties which differ substantially from the singular isothermal
profile \cite{li00,perrotta01}.

For this purpose, we use a sample of five clusters obtained from dark
matter $N$-body simulations, kindly made available by the GIF
collaboration. These are placed at redshift $z_{\rm L}=0.27$ and have
virial masses ranging between $\sim 3\times 10^{14} M_\odot/h$ and
$\sim 10^{15} M_\odot/h$. The cosmological model is the $\Lambda$CDM
model ($\Omega_{\rm M}=0.3$, $\Omega_\Lambda=0.7$, $h=0.7$,
$\sigma_8=0.8$). A complete description of the numerical cluster
sample can be found in Bartelmann et al.~\cite{bartelmann98}. We use
that sample to perform ray-tracing simulations
\cite{meneghetti00,meneghetti01}, placing a large number of elliptical
sources at redshift $z_{\rm S}=1$. Finally, we measure the lensing
cross sections for arcs with length-to-width ratio larger than $7.5$.

\begin{figure}
  \plotone{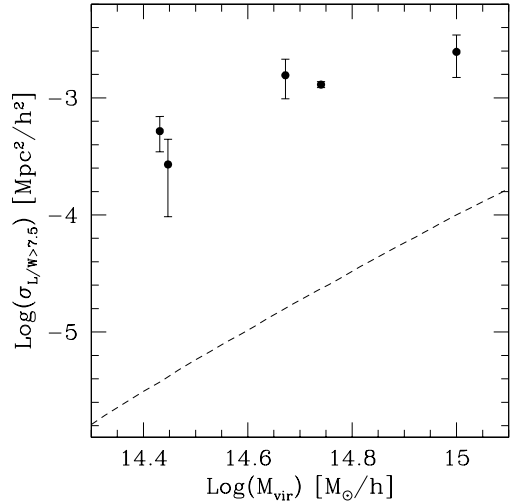}{7.0cm}
\caption{Comparison between numerical and spherical analytic models:
the dashed line indicates the cross section for arcs with
length-to-width ratio $>7.5$ of NFW spheres as a function of the lens
mass; filled circles indicate the cross sections of the five numerical
clusters used in the ray-tracing simulations.}
\label{figure:figure1}
\end{figure}

The results are shown in Fig.~\ref{figure:figure1}. The plot displays
the numerically measured cross sections (filled circles), obtained by
averaging the values corresponding to three independent projections of
each cluster, as a function of the virial mass of the lens. The error
bars show their uncertainties. The dashed line indicates the
analytically computed lensing cross section of NFW spheres placed at
the same redshift of the simulated clusters. The numerical cross
sections are systematically and substantially larger than the analytic
ones by roughly two orders of magnitude. This clearly shows that
spherically symmetric models, even with a very realistic density
profile, fail to reproduce the strong-lensing properties of
numerically modeled galaxy clusters.

A possible explanation for this discrepancy is the lack of asymmetries
and substructures in the matter distribution of the spherical
models. In fact, the importance of both these factors for the strong
lensing efficiency of galaxy clusters was demonstrated by Bartelmann
et al.~\cite{bartelmann95}.

\section{Elliptical NFW lensing potential}

The lack of asymmetries and substructures in the matter distribution
could be compensated by adding ellipticity to the lensing potential of
the NFW spheres.

The spherical NFW lensing potential is given by
\begin{equation}
  \Psi(x) \propto \left(\frac{1}{2}\ln^2\frac{x}{2}-
  2\mbox{arctanh}^2{\sqrt{\frac{1-x}{1+x}}}\right)\;,
\end{equation} 
where $x=r/r_{\rm s}$ is the distance from the lens centre $r$ divided
by the scale radius $r_{\rm s}$ of the dark-matter halo.

We introduce the ellipticity $e\equiv 1-b/a$ by substituting
\begin{equation}
  x \rightarrow \rho=\sqrt{\frac{x_1^2}{(1-e)}+x_2^2(1-e)}\;.
\end{equation}
The components of the deflection angle are then
\begin{eqnarray}
  \alpha_1 = \frac{\partial\Psi}{\partial x_1} =
  \frac{x_1}{(1-e)\rho}\hat{\alpha}(\rho) & ; &
  \alpha_2 = \frac{\partial\Psi}{\partial x_2} =
  \frac{x_2(1-e)}{\rho}\hat{\alpha}(\rho)\;,
\label{equation:angles}
\end{eqnarray}
where $\hat{\alpha}(\rho)$ is the unperturbed (i.e. spherical)
deflection angle at the distance $\rho$ from the lens centre.

\begin{figure}
  \plottwo{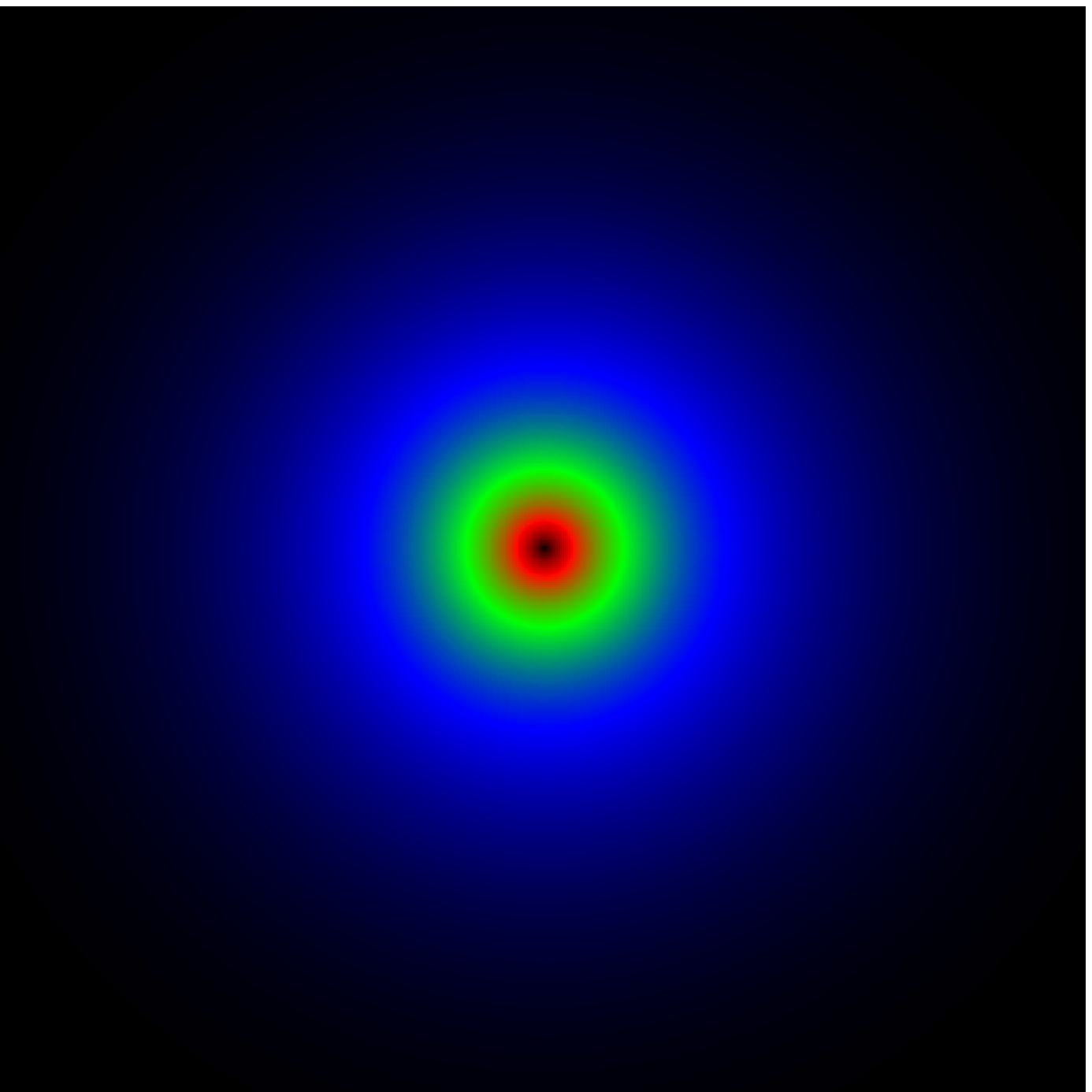}{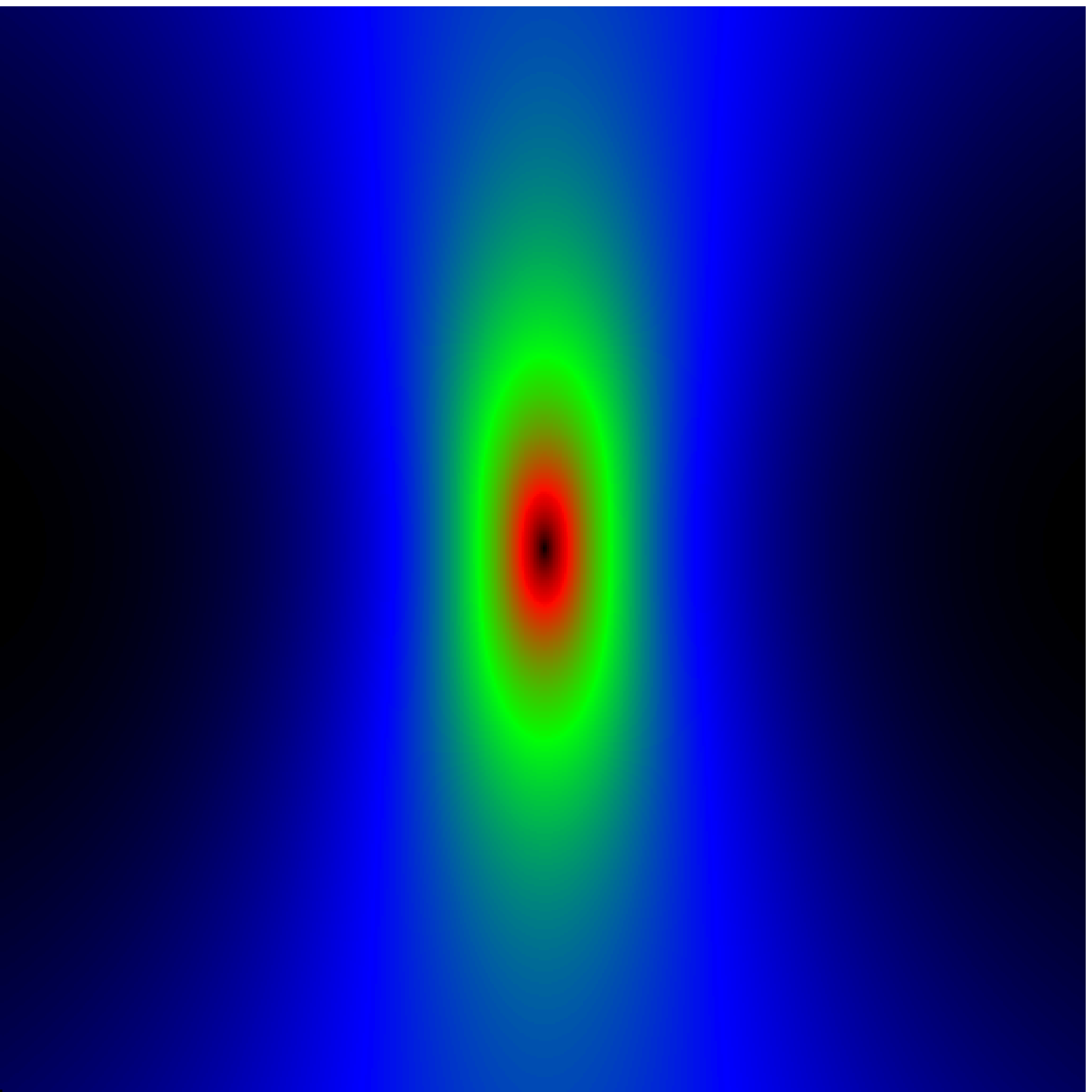}
\caption{Deflection-angle maps obtained assuming a spherically
symmetric NFW lensing potential (left panel), and after adding an
ellipticity of $e=0.3$ (right panel), as explained in the text.}
\label{figure:maps}
\end{figure}

\begin{figure}
  \plotone{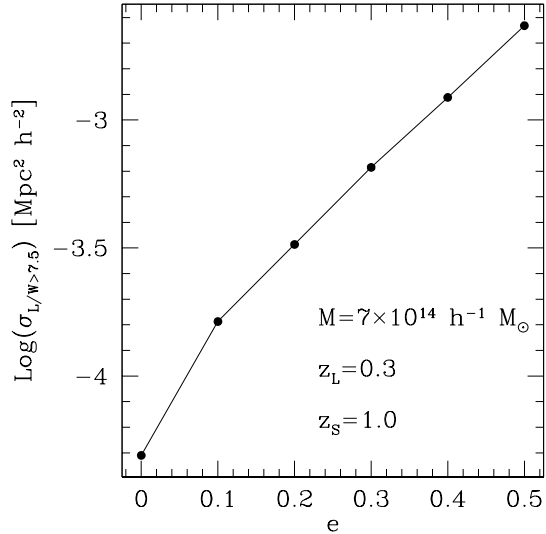}{7.5cm}
\caption{Lensing cross section for arcs with length-to-width ratio
larger than $7.5$ as a function of the ellipticity added to the
lensing potential. The lens mass is $7 \times 10^{14} M_\odot/h$,
and its redshift is $z_{\rm L}=0.3$.}
\label{figure:sigmae}
\end{figure}

Using (\ref{equation:angles}), we produce deflection angle maps like
those displayed in Fig.~\ref{figure:maps}. The left panel of the
figure shows the unperturbed deflection angle map for a lens of $7.5
\times 10^{14} M_\odot/h$ at redshift $z_{\rm L}=0.3$. The right panel
shows the map obtained by adding an ellipticity of $e=0.3$.  Using
these deflection-angle maps in the ray-tracing simulations, we measure
the lensing cross sections of the elliptical models. These are
displayed in Fig.~\ref{figure:sigmae} as a function of the ellipticity
added. The results show that an ellipticity of $e=0.3$--$0.4$ could
suffice to remove the discrepancy between the numerical and the
analytic models. Therefore, the usage of elliptical models is more
appropriate for computing the number of arcs with a minimal
length-to-width ratio produced by a sample of galaxy clusters via
gravitational lensing.

\section{Conclusions}

Upcoming full-sky microwave surveys will detect several thousands of
galaxy clusters through their thermal SZ effect. Many of these
clusters will also be efficient gravitational lenses.

We can predict how many {\em giant\/} arcs will be produced by this
cluster sample by using analytic models to describe the lenses. We
verified that spherical models, even with realistic profiles, do not
reproduce the lensing properties of numerically modeled galaxy
clusters: We need to include at least ellipticity in order to give a
correct description of the cluster lenses.

We now plan to calibrate the analytic lensing cross sections for the
asymmetric NFW halo profile with numerical cluster models. This can be
done by measuring the ellipticity of the lensing potential of the
simulated clusters by fitting their deflection angle maps to those of
perturbed analytic models.

\vfill
\end{document}